# Towards an Explanatory and Computational Theory of Scientific Discovery

*Chaomei Chen*[1,2]*, Yue Chen*[2]*, Mark Horowitz*[1]*, Haiyan Hou*[2]*, Zeyuan Liu*[2]*, Donald Pellegrino*[1]
[1]College of Information Science and Technology, Drexel University, USA
[2]The WISE Lab, Dalian University of Technology, China
Email: chaomei.chen@cis.drexel.edu

**Abstract**

We propose an explanatory and computational theory of transformative discoveries in science. The theory is derived from a recurring theme found in a diverse range of scientific change, scientific discovery, and knowledge diffusion theories in philosophy of science, sociology of science, social network analysis, and information science. The theory extends the concept of structural holes from social networks to a broader range of associative networks found in science studies, especially including networks that reflect underlying intellectual structures such as co-citation networks and collaboration networks. The central premise is that connecting otherwise disparate patches of knowledge is a valuable mechanism of creative thinking in general and transformative scientific discovery in particular. In addition, the premise consistently explains the value of connecting people from different disciplinary specialties. The theory not only explains the nature of transformative discoveries in terms of the brokerage mechanism but also characterizes the subsequent diffusion process as optimal information foraging in a problem space. Complementary to epidemiological models of diffusion, foraging-based conceptualizations offer a unified framework for arriving at insightful discoveries and optimizing subsequent pathways of search in a problem space. Structural and temporal properties of potentially high-impact scientific discoveries are derived from the theory to characterize the emergence and evolution of intellectual networks of a field. Two Nobel Prize winning discoveries, the discovery of *Helicobacter pylori* and gene targeting techniques, and a discovery in string theory demonstrated such properties. Connections to and differences from existing approaches are discussed. The primary value of the theory is that it provides not only a computational model of intellectual growth, but also concrete and constructive explanations of where one may find insightful inspirations for transformative scientific discoveries.

**Keywords**: Theory of scientific discovery, transformative scientific discoveries, theory of structural holes, intellectual brokerage, knowledge diffusion, information foraging

# 1 Introduction

The intellectual structure of a scientific field is an abstraction of the collective knowledge of scientists in the field, including scholarly publications and other forms of intellectual assets. *Scientific change* refers to profound changes of the intellectual structure of a field. In this article, we will focus on the nature and key mechanisms of scientific discoveries that could lead to such fundamental changes – *transformative scientific discoveries*.

The nature of scientific change has been studied from many distinct perspectives, notably including philosophy of science (Collins, 1998; Laudan et al., 1986; Schaffner, 1992), sociology (Fuchs, 1993; Griffith & Mullins, 1977), and history of science (Brannigan & Wanner, 1983). Quantitative studies of the topic can be found in the fields of scientometrics, citation analysis, and information science in general (Chen, 2003; Heinze & Bauer, 2007; Heinze, Shapira, Senker, & Kuhlmann, 2007; Hummon & Doreian, 1989; Small & Crane, 1979; Sullivan, Koester, White, & Kern, 1980; Wagner-Dobler, 1999). Scientific literature has increasingly become one of the most essential sources for these studies. Social network analysis and complex network analysis also provides valuable perspective (Barabási et al., 2002; Newman, 2001; Redner, 2004; Snijders, 2001; Valente, 1996; Wasserman & Faust, 1994).

What do these diverse perspectives have in common and how do they differ in terms of their views of scientific change, scientific discovery, and knowledge diffusion? In this article, we introduce an explanatory and computational theory of scientific discovery as a key step for understanding and explaining the emergence and evolution of the intellectual structure of a field. We are motivated for a number of reasons. First, despite of the perceived role of serendipity and other unpredictable factors, it is evident that an important subset of scientific discoveries share important and generic properties (Bradshaw, Langley, & Simon, 1983; H. A. Simon, P.W. Langley, & G. L. Bradshaw, 1981). In order to obtain conclusive evidence, one will need a theory of scientific discovery that can provide a unifying conceptual framework so that one can characterize a variety of scientific discoveries from a consistent perspective. Second, given one concrete case of scientific discovery, it may be studied from multiple and often not interconnected perspectives. For example, a philosophical study of a scientific revolution may have little overlap with a sociological study of the same process. Even two philosophical studies of the same scientific revolution could appear to be unrelated in the eyes of laypersons. We need a theory that can not only explain scientific change, but also relate to various existing theories. Third, statistical models of network evolution have been used to identify statistical and topological properties of scientific networks. However, such properties, although generic in nature, do not readily offer further explanations of why scientists in a network behave in a particular way. Motivations, decisions, and interpretations underlying such properties are often detached or left out. Thus, we need a theory that not only identifies statistical and topological properties of scientific networks, but also offers practical insights into the mechanisms that may drive scientists' observed behavioral patterns. The work described in this article is the first step towards this long-term goal.

There are many types of theories, including descriptive, explanatory, generative, predictive, and prescriptive (Bederson & Shneiderman, 2003). Our immediate goal is to develop a simple, descriptive, explanatory, and generative theory of scientific discovery. We are interested in identifying some generic mechanisms of discovery in order to explain transformative scientific discoveries to begin with and other types of discoveries later on. Such generic mechanisms are in fact generative in nature because scientists and computer simulation algorithms would be able to emulate such mechanisms. We have a few expectations of our new theory. First, it should help us to recognize the significance of new discoveries as soon as possible. Second, it should help us to identify as many potential areas of growth as possible. Third, it should help us to explain both the creation of knowledge and its diffusion within a consistent and unified framework.

The rest of the article is organized as follows. We will first review existing conceptualizations of scientific change in the philosophy of science, sociological theories of scientific change, sociological theories of creative ideas, information foraging theory, and a recurring theme among these various views. The recurring theme is, simply speaking, that insights, creative ideas, and transformative scientific discoveries are the work of a broad range of brokerage mechanisms.

Next, we will expand the recurring theme and construct a simple theory of scientific discovery to explain the growth of a scientific field. We will then describe conjectures that one can derive from the first principles of the theory, including structural and temporal properties of citation and co-citation networks. We will include a brief analysis of Nobel Prize winning discoveries as illustrative cases. Finally, we will outline ongoing and future work, including large-scale computer simulation and a wider range of high-impact scientific discoveries.

# 2 Existing Conceptualizations of Scientific Change

## 2.1 Specialties and Scientific Change

*Specialty* is a key concept in the study of scientific change. A specialty is a group of researchers and practitioners who have similar training, attend the same conferences, read and cite the same body of literature (Fuchs, 1993). There are a variety of studies of specialty in the literature (Chubin, 1976; Fuchs, 1993; Morris & Van der Veer Martens, 2008; Mullins, Hargens, Hecht, & Kick, 1977; Small & Crane, 1979). For example, Mullins et al. studied author groups corresponding to co-citation clusters using questionnaires and concluded that co-citation clusters indeed represent the intellectual structure and that coauthors do form social groups (Mullins et al., 1977). Co-author networks have also been studied in complex network analysis of community structures (Girvan & Newman, 2002). These finding provide an empirical basis for the analysis of scientific change based on co-citation networks as we shall introduce later in this article.

The dynamics of the structure of a specialty is a central issue in the context of scientific change. Research has shown that major changes in a variety of disciplines tend to be originated within small, socially coherent groups (Griffith & Mullins, 1977). Kuhn observed that new paradigms are typically initiated by young scientists or newcomers to a crisis-laden field (Kuhn, 1962). In addition, Crane (1969) found that the desire for originality motivates scientists to maintain contacts with scientists and scientific work in areas *different from their own* in order to enhance their ability to develop new ideas in their own areas. This observation underlines an intriguing fact that many major scientific discoveries are often fundamentally inspired by external influences, or from peripheral areas of established research specialties, which echoes Kuhn's earlier observation.

Crane's observation can be seen as a special case of what sociologist Burt called the social capital of *structural holes* (Burt, 1992, 2001, 2004). Structural holes are voids in social structure. According to Burt's theory of structural hole, structural holes in a social network are disconnected or poorly connected areas between tightly and densely connected groups of people. The presence of such structural holes may influence the importance of positions in a social network – some positions become more privileged and competitive than others. The value of a person in a social network is therefore linked to the potential to establish connections between groups that are separated by structural holes. People in positions with great brokerage potentials are known as brokers and gatekeepers. Brokers are rewarded for their integrative work in terms of more positive evaluations, higher compensations, and faster promotion. The underlying reason for the difference is that information is more homogeneous within groups, whereas more heterogeneous between groups. Brokers are in special positions to access heterogeneous information from a broader range of sources. The privilege often leads to competitive advantage. In the following sections, we will argue that the role of brokerage mechanisms not only goes beyond social networks, but also underlines an important source of insight that leads to profound scientific changes and discoveries.

The dynamics of theory-change in science is not only a philosophical issue, but also a historical one. Brush investigated whether scientists give greater weight to novel predictions than to explanations of known facts against historical cases in physical science (Brush, 1994). Several theories were accepted after successful novel predictions but there is little evidence that extra credit was given for novelty. Others were rejected despite, or accepted without, making successful novel predictions. No examples were found of theories that were accepted primarily because of successful novel predictions and would not have been accepted if those facts had been previously known. Brush further examined the impact of predictions on theory acceptance through several cases, including the Big Bang vs. steady-state cosmology, the origin of the Moon, gravitational light bending, and Hannes Alven's plasma physics (Brush, 1995). Brush concluded that confirmed predictions provide "corroboration" of a hypothesis, but only in the minimalist sense of scientists voting with their publications. Corroboration "merely makes it more reasonable to pursue that hypothesis than one that has not been corroborated," and thus "there was a significant increase in publications on the theory [i.e., those theories in the case studies] that led to the prediction" (p. 314). Brush also emphasized, however, that "if one's basic assumptions and method are considered unacceptable by other scientists, no amount of empirical confirmation will force them to accept it. It is said not as a criticism of the scientific community, but simply as a fact about science which many philosophers of science ignore" (p. 307).

A mathematical approach to the prediction of scientific discovery was proposed in (William Goffman & Harmon, 1971). Their approach is built on a four-state Markov chain model of discovery. Discovery is conceptualized as a process of placing a set of information in the right order. They were able to construct such a model based on an expert-annotated bibliography of the field of symbolic logic. The discovery per se would be the ordered information, i.e. patterns. The four states are defined in terms of the sufficiency and order of information. In State I, information is insufficient and unordered. The problem at this stage is to acquire information, not to order it. Observations are inadequate to establish patterns. In State II, information is insufficient but available information is ordered. In State III, there are sufficient information elements, but not in the right order. Finally, in state IV, information is both sufficient and ordered. The discovery is established. From here, it can be elaborated, refined, or challenged.

## *2.2 Knowledge Diffusion*

Knowledge diffusion, or the spread of knowledge, is an important aspect of the dynamics of a specialty. As we will demonstrate in this article, knowledge diffusion can be explained as an information foraging process stimulated by the original scientific discovery.

### 2.2.1 Quantitative Models of Diffusion

Quantitative models of how scientific ideas spread are proposed by many researchers (Bettencourt, Castillo-Chavez, Kaiser, & Wojick, 2006; Bettencourt, Kaiser, Kaur, Castillo-Chavez, & Wojick, 2008). Epidemic models are among the most popular ones (W. Goffman & Newill, 1964; Liben-Nowell & Kleinberg, 2008; Nowakowska, 1973). Epidemic models consider variables such as contact rates between scientists, latency and recovery times. The contact rate between scientists is found to be the single important factor to speed up the diffusion of knowledge.

Other potentially applicable models of diffusion include ant colony and random walk models. In an ant colony model (Dorigo & Gambardella, 1997), ants travel between their home and food sources. They leave scents as trails for others. Scents decrease over time unless being reinforced by other ants. One can see a natural mapping from the ant colony model to a model of network evolution. Ants are replaced by scientists. Their home is now the contemporary intellectual structure. The food sources are new publications in the literature. Finding foods is equivalent to making a reference to a new publication. Doing so also leaves trails for other scientists. Ant colony is a self-organizing optimization mechanism. Unlike the preferential attachment approach, which specifies the criteria of an addition to an existing network, ant colony is not limited to preferences, although it can be tailored to make use of them.

Random walk algorithms are also useful for modeling the spread of information. A random walk model over a network is built on state transition probabilities. Each node in the network represents a state. Moving from one node to another is governed by a state transition probability, which can be updated based on available evidence in Bayesian rules. The spread of knowledge is thus translated into a question of how easy or how hard it would be to make such moves.

The ant colony and random walk models have a more profound connection to the *information foraging theory* (Pirolli, 2007). The fundamental premise of the information foraging theory is that the behavior of a forager, namely, information searchers and, in this case, scientists, is driven by a perceived or calculated profitability of the potential move. The profitability takes into account the expected returns as well as potential risks or costs involved. For example, if online access to an article costs $30, then the cost is only part of the equation. Whether the article is worth your paying the $30 depends on what you can do with the article and how urgently you need it.

Sandstrom argued that information seekers are very much like foragers, especially in terms of how and where they forage for valued resources (Sandstrom, 1999). She introduced the notion of bibliographic microhabitats to underline the similarity between hunters and information seekers. She further argued that if some empirical cost-benefit currency can be established, then analysts would be able to rank foragers' preferences, predict which resources will be pursued, and specify the net returns associated with particular choices.

In summary, unlike epidemic models, foraging models emphasize not only structural properties of an information space for information seekers or a problem space for scientists, but also the interplay between perceived values, handling costs, and various competing and probably conflicting factors in a broader context of decision making. In other words, one may incorporate foraging models into existing workflows so that one can recognize and act upon vital clues that may lead them to a fruitful path.

### 2.2.2 Searching for Indicators of Great Discoveries

What is the extent to which quantitative rankings of highly cited authors confirm or, even more ambitiously, predict Nobel Prize awards? Between 1977 and 1992, Garfield published a series of studies of Nobel Prize winners' publications and their citations and made predictions of future Nobel Prize laureates based on existing citation data. He reported that eight Nobel laureates were found on a list of 100 most cited authors from 1981 through 1990(Garfield & Welljamsdorof, 1992). Others on the list were seen as potential Nobel Prize winners in the future. On the other hand, it was noted that the undifferentiated rankings of the most cited authors in a given period of time could be further fine-tuned to increase the accuracy of its coverage of Nobel Prize awards.

For example, the Nobel Committee sometimes selects relatively small specialties. Further dividing the list according to specialties shows that Nobel laureates in relatively small specialties are among the most cited authors in their specialties.

Methods papers of Nobel Prize winners tend to attract a disproportionably high amount of citations. More recent examples of methodological contributions include the 2007 Nobel Prize for the British embryonic stem cell research architect Martin Evans. Garfield coined the phenomenon the *Lowry Phenomenon*, referring to the classic example of Oliver Lowry's 1951 methods paper, which was cited 205,000 times up to 1990.

Research has shown that citation frequency has a low predictive power for Nobel awards because there are so many other scientists with the same or even higher citations as the few Nobel Prize winners. The greatest value of counting citations is its simplicity. Subsequent attempts to improve the accuracy of the method tend to lose the simplicity. Hirsch's h-index has drawn much interest also because of its simplicity despite its known limitations (Hirsch, 2005a). Antonakis and Lalive intended to capture both the quality and productivity of a scholar with a new index IQp (Antonakis & Lalive, 2008). They compared the new index of Nobel winners in physics, chemistry, medicine, and economics. It is worth noting here that one should always be cautious when using quantitative indicators in qualitative decisions. The authors found about two third of Nobel winners have an IQp over 60. The authors showed that in several examples, IQp differentiated Nobel class and others more accurately than the h-index, including physicist Ed Witten (h=115 and IQp=230) and others who have high h-index but relatively low IQp index, S. H. Snyder (h=198, IQp=117) and R. C. Gallo (h=155, IQp=75).

Börner and her coauthors proposed a co-evolution model of networks of authors and networks of papers based on growth mechanisms such as preferential attachment (Börner, Maru, & Goldstone, 2004). They validated their model with real-world data from the *Proceedings of the National Academy of Science of the United States of America*. Small studied tracking and predicting growth areas in science based on co-citation clusters and relative ages of clusters (Small, 2006). We will introduce a different growth mechanism in our theory. Our mechanism provides an alternative approach to the preferential attachment one.

In the context of scientific discovery, we will expand the information foraging theory to describe the behavior of scientists in searching for novel hypotheses and theories. This will help us to explain how a scientist would maximize the profitability of the next move.

## *2.3 Common Mechanisms for Scientific Discovery*

There is evidence in the literature that scientific discoveries do share some common mechanisms, especially in light of research in computer simulation of discoveries, cognitive studies of scientific change, and the nature of insight.

### 2.3.1 Scientific Discovery as Problem Solving

The most prominent work in this area has been done by Herbert Simon and his colleagues using computer simulation to study and reconstruct scientific discoveries (Bradshaw et al., 1983). A long list of examples of automated discoveries was given in (Glymour, 2004). He used the metaphor of finding a needle in a haystack to characterize the problems faced by scientists in discovery. Rather than sifting through things in the haystack one by one, automated discovery is akin to either setting the haystack on fire and blowing away the ashes to find the needle, or running a magnet through the haystack. There are advantages and limitations. Following the metaphor, for example, the fire may melt the needle.

Many studies have addressed the nature of insight in scientific discovery. For example, Gestalt psychologists suggest that insight occurs when problem solvers see the original problem from a fresh perspective (Mayer, 1995). Other researchers have emphasized that the complexity of searching in a problem space has more to do with the structure of a problem space than the searcher (Perkins, 1995; Simon, 1981). In particular, Perkins distinguished two types of problem spaces. In a Homing Space, there are many clues and signposts such that navigating in such spaces is relatively easy. In contrast, a Klondike Space has very few such clues. The sparseness of clues is illustrated by Perkins (p. 498) in a widely known case of sudden insight – Charles Darwin's discovery of the principle of natural selection. According to Darwin's autobiography, in October 1838, he conceived the principle while he "happened to read for amusement 'Malthus on Population.' What is remarkable is that the next person arrived at the same principle 20 years later. What is even more remarkable is that the person, Alfred Russell Wallace, arrived to the idea while reading the same 1826 book by Malthus!

How could one increase the odds of stumbling on such clues? It becomes clear, from Sandstrom's notion of bibliographic microhabitats to Perkins' characterizations of Homing and Klondike spaces, that finding and recognizing clues is essential for both information foragers and problem solvers. Research in the data mining community on interestingness is particularly relevant (Hilderman & Hamilton, 2001; Liqiang & Howard, 2006). Interestingness is a quantitative measure of where a set of scientific ideas fit on the spectrum which ranges from the practice of normal science to that of paradigm-shifting ideas (Davis, 1971). In this regard, interestingness lies between order and complete randomness, or chaos. We posit that three distinct ranges of scientific reports and ideas are those which are (1) either confirmatory or boring – there is nothing new for the scientific reader; the previously stated hypotheses have not been falsified yet, and are less and less likely to be so determined; (2) the interesting ideas or work, which denies widely accepted assumptions, states new relationships between old ideas, proposes new mechanisms, but do not require the reader to adopt wholly new ways of thinking; and (3) paradigm shifts and transformative discoveries. Interesting ideas are enlightening and surprising in a non-threatening way; in fact, the surprise is generally a pleasant one, in contrast to the experience of living through a shift of paradigm, especially when one's accepted paradigm is being replaced by a more successful one.

### 2.3.2 Literature-Based Discovery

Swanson and his colleagues pioneered a literature-based discovery approach to identify potentially valuable hypotheses (Swanson, 1986a, 1986b; Swanson, 1987; Swanson & Smalheiser, 1999). In essence, according to Swanson, the model of discovery from public knowledge is the A-B-C model, where the connections of A-B and B-C are known, but the connection of A-C is unknown. Thus A-C has the potential to become a candidate hypothesis for domain experts to evaluate. Using this template, a series of such candidate associations have been identified, including the connections between fish oil and Raynaud's syndrome (Swanson, 1986a), magnesium and migraine (Swanson, 1988), indomethacin and Alzheimer's disease (Smalheiser & Swanson, 1996).

Many researchers have subsequently adapted and refined Swanson's techniques. For example, Gordon and Lindsay conducted experiments with the MEDLINE medical literature database and extended the work of Swanson (Gordon & Lindsay, 1996; Lindsay & Gordon, 1999). They used lexical statistics to discover hidden connections in the medical literature. They argued that hidden connections are those that are unlikely to be found by examination of bibliographic citations or the use of standard indexing methods and yet establish a relationship between topics that might profitably be explored by scientific research. They also mentioned that literature-based discovery cannot replace traditional empirical scientific research or even literature search, but rather supports them by providing the scientist with a means to organize more easily a potentially overwhelming amount of information.

Recently, Kostoff and his colleagues published a series of studies of literature-based discovery. These special studies presented a comprehensive approach for systematic acceleration of discovery and innovation, and demonstrated the generation of large amounts of potential discovery through five case studies describing the application of literature-based discovery to Raynaud's syndromes, cataracts, Parkinson's disease, multiple sclerosis, and water purification. He described the lessons learned from each application, and how the techniques can be improved further (Kostoff, 2008).

Where can we go from here? How often could a Nobel Prize award be characterized in terms of this A-B-C pattern of transitivity? Are there other patterns of scientific discoveries? If literature-based discovery is a computer-aided search in a problem space, what would it miss?

### 2.3.3 Thinking Outside the Box

Effective strategies for making scientific discoveries have highlighted the ability to think creatively and look at a problem from a fresh perspective. Dunbar, for example, compared two different strategies of hypothesis generation using a Nobel Prize winning discovery as the test case (Dunbar, 1993). He found that it is a more effective discovery strategy to encourage researchers to consider novel alternative hypotheses. A 1992 special issue of *Theoretical Medicine* examined the mechanisms of scientific revolution and how the Nobel Prize committee selected scientific discoveries (Lindahal, 1992).

A longitudinal study of highly creative scientists in nano science and technology has found that it is not only the sheer quantity of publications that enables scientists to produce creative work but also their ability to effectively communicate with otherwise disconnected peers and to address a broader work spectrum (Heinze & Bauer, 2007). Why is it possible that communicating with otherwise disconnected scientists can lead to more creative work? What can one do specifically to come up with novel alternative hypotheses? How do we think outside the box?

## *2.4 Connecting Diverse Perspectives*

There are many philosophical theories of scientific change. Philosophers of science (Laudan et al., 1986) argue that it would be useful to compare rival theories of scientific change against the history of science. Proponents suggest that conjectures of philosophical theories should be organized into theses so that one can compare these theories in terms of individual theses. Laudan et al. recommended rephrasing Lakatos' research programme, Laudan's research tradition, and Kuhn's paradigm in terms of a more generic notion of guiding assumptions. A superior theory of scientific change would be the one that has the most matches from the historical data. This idea was later criticized by (Radder, 1997), suggesting that it was far too ambitious.

Our needs here are different. Our goal is not to evaluate the value of individual philosophical theories of scientific change. Rather, what we need is an explanatory theory that can clarify the underlying mechanisms of specific scientific discoveries. In addition, we need a theory that can be instrumental for quantitative studies of scientific change.

Kuhn's paradigm-shift model of scientific revolutions (Kuhn, 1962, 1970) is probably the most widely known theory. It describes how science advances through a path of normal science, crisis, revolution, and new normal science. A revolution involves a shift of world views from an old paradigm to a new paradigm. The paradigm-shift model has drawn criticism. Critics argue that scientific change is often a lengthy process instead of a swift change as the paradigm-shift model suggests.

Cognitive scientists consider scientific discovery in common with everyday problem solving (Herbert A. Simon, Patrick W. Langley, & Gary L. Bradshaw, 1981). In (Klahr & Simon, 1999), four approaches to research on scientific discovery were identified; namely, historical accounts of scientific discoveries, psychological experiments with nonscientists working on tasks related to scientific discoveries, direct observation of ongoing scientific laboratories, and computational modeling of scientific discovery processes—by viewing them through the lens of the theory of human problem solving. The authors then considered these types of studies against a list of evaluative criteria, such as face validity, fine or coarse-grained, new phenomena, rigor and precision, social and motivational factors.

Many scholars have studied information and discovery pathways. Small presented a series of examples from the history of science in which discoveries can be modeled as navigation between pairs of established experimental or theoretical findings (Small, 2000). One of his examples was from atomic physics in early twentieth century. There was no direct connection between experimental evidence on the spectrum for atomic hydrogen and evidence for hydrogen's nuclear structure until Niels Bohr's 1913 model for the hydrogen atom using a quantum hypothesis. Similarly, the Müller-and-Bednorz discovery of superconductivity was also seen as creating a path between the field of superconductivity and a class of compounds previously not thought to be promising candidates for superconductivity (Holton, Chang, & Jurkowitz, 1996; Small, 2000).

We notice a recurring theme in the diverse conceptualizations of scientific change. That is, profound scientific change tends to be connected to a broad range of brokerage mechanisms. Burt's structural holes are found not only in social networks but also in associative networks of intellectual, semantic, and other types of interrelationships. Because information flow around a structural hole is limited by the topological structure, those who are in the brokerage positions inherit advantages from their positions in such networks. Furthermore, structural holes in intellectual and cognitive networks appear to be a vital source of inspiration and creativity. Creative scientists draw inspirations from other disciplines. Research has found that great philosophers tend to be the ones who stayed in touch with competing schools of philosophy (Guiffre, 1999). Creative scientists are the ones who have the ability to communicate effectively with otherwise disconnected peers (Heinze & Bauer, 2007). Scientists make extra efforts to maintain contacts with scientists in different fields (Crane, 1972). Therefore, we have reached our central premise: bridging structural holes in a knowledge space is a valuable and viable mechanism for understanding and arriving at transformative scientific discoveries.

## 2.5 Bridging Intellectual Structural Holes

Now we will review some of the major conceptualizations of scientific change in light of the theory of structural holes (Burt, 1992, 2004, 2005). The theory of structural holes was originally developed in the context of social networks. We will show that the theory provides a meaningful and indeed enlightening framework for explaining why structural holes in intellectual networks such as co-citation networks may play an essential role in scientific discovery. Although this new conceptualization goes beyond the original scope of Burt's theory, we still refer them as structural holes for simplicity.

According to a sociological theory of scientific change (Fuchs, 1993), scientific discoveries are driven by two social factors, namely, peer competition and mutual dependence. Scientists seek novel discoveries to stay ahead in the invisible competition with their peers. As we have learned from the large body of relevant literature, inspirations often rise when different ways of thinking interact with one another. Structural holes in this sense span across patches of knowledge at different levels of self-organized structures, ranging from areas of research, fields of study, to disciplines.

From the information foraging perspective, establishing conceptual linkages between disparate patches of knowledge is a high-risk and high-return action. On the one hand, adapting a theory or a method from a 'foreign' discipline is likely to ensure its novelty in the 'home' discipline. It is more likely to think 'outside the box' in such situations. On the other hand, the fact that ideas and inspirations have obviously worked in another domain will drastically reduce the potential risk that scientists may have to bear. This combination appears to give the maximum profitability associated with a structural hole.

From a philosophy of science's point of view, focusing on a structural hole also makes sense. In terms of Kuhn's paradigm-shift model, a competing paradigm is more likely to come from an unexpected place than right from the center of the currently predominant paradigm. We will give detailed descriptions of the theory of scientific discovery in the following section.

# 3 An Explanatory and Computational Theory of Discovery

A recurring theme across a wide variety of studies of scientific discovery, scientific change, creativity, and insight is that many creative ideas and profound discoveries can be traced to the work of a generic class of brokerage mechanisms. Brokerage mechanisms are not only found in social networks, such as networks of collaborators and coauthors, but also in the more abstract conceptual networks of scientific knowledge, such as co-citation networks. For example, brokerage mechanisms have been seen to establish a previously unexpected linkage between structures of knowledge, connect two or more previously disparate fields of study, or recognize a meaningful analogy between distinct theories or hypotheses. Our new theory of scientific discovery is built on this recurring theme.

## 3.1 Basic Elements of the Theory

As the first step towards an explanatory and computational theory of scientific discovery, we will focus on transformative and revolutionary discoveries. *Transformative discoveries* represent fundamental and revolutionary scientific changes. The growing interest in cyber-enabled discovery, e-science, and e-social science underlines the importance of advancing our understanding of how science works and identifying recurring mechanisms of creativity and discovery (Shneiderman, 2002, 2007). Supporting more transformative research is of critical importance in the fast-paced, science and technology-intensive world (NSF, 2007).

The fundamental premise of our theory is that *a transformative discovery is made when a novel connection is established between two or more previously disparate units of scientific knowledge.* Disparate units of scientific knowledge may include unconnected theories in different disciplines, isolated observations in the same field, or publications that have never been thought to be related. This conception is related to a number of approaches in the literature.

First, the brokerage-focused theory is inspired by the structural-hole theory of social networks (Burt, 2005). Furthermore, our theory adapts the brokerage mechanism and introduces it as a generic discovery mechanism for a wide variety of networks of scientific knowledge, such as citation networks, co-citation networks, networks of collaborating scientists, and other associative networks. The hypothesis that brokerage leads to greater collaborative creativity was tested in a recent study of collaborative inventors of utility patents (Fleming, Mingo, & Chen, 2007). Fleming et al. demonstrated that cohesive networks hamper creativity but aid in its transfer, particularly if the knowledge is complex and tacit. They tested more specific hypotheses such as a person is more likely to create new combinations if he or she brokers relations between otherwise disconnected collaborators. New combinations, as integrative work, are defined as a mechanism of creativity. In contrast, our theory focuses on transformative discoveries, which are conceptually more complex than new combinations of existing discoveries. For example, transformative discoveries often introduce new concepts and theories before integrative work becomes possible. The brokerage view also provides a simple explanation of why communicating with otherwise disconnected peer scientists is a distinct character of creative scientists (Heinze & Bauer, 2007).

Second, our theory is also related to literature-based discovery in that it shares the general goal of finding generative mechanisms of discovery. On the other hand, it differs from Swanson's famous A-B-C model. Instead of searching for a transitive closure of A➔C, given A➔B and B➔C, we focus on the brokerage mechanism of discovery, which aims to establish an innovative connection between A and C. Another important difference is that we utilize structural properties of a network, whereas such properties are not used in Swanson's approach and its variations.

Third, our theory is related to network evolution models in complex network analysis. Preferential attachment models, for example, characterize the growth of a network in a process that popular nodes will become even more popular as new nodes and links are added to the network (Albert & Barabasi, 2002; Barabási et al., 2002). The popularity of a node can be broadly defined by an attribute function of node, such as prestige, age, or by other ranking mechanisms. Such processes often result in scale-free networks, which are characterized by power law distributions of node degrees. While earlier preferential attachment models assume that each new coming node is fully aware of the prestigious status of every existing node, more recent studies have relaxed the assumption to ranking functions defined on a subset of the existing nodes instead (Fortunato, Flammini, & Menczer, 2006). In contrast, the brokerage mechanism in our theory provides a growth mechanism by building connections across structural holes between two or more thematic networks. A brokerage-driven growth is distinct from growths that can be modeled by preferential attachment.

Fourth, our theory extends earlier efforts for predicting Nobel Prize winners based on citation ranking (Garfield, 1992). Thomson Reuters' Citation Laureates[1] are also in this category. Our approach is distinct in several important ways. Although using citation ranking alone has the advantage of simplicity, we take multiple factors such as structural holes and the rate of citation growth into account in order to better accommodate the complexity. In addition, we are concerned with the possibly delayed identification due to the time taken for the citation profile of a scholarly publication to become prominent enough to be noticed. We expect that using structural properties in the theory can resolve the issue to some extent.

[1] http://scientific.thomsonreuters.com/nobel/nominees/

Fifth, our theory provides an explanatory mechanism for the diffusion process associated with a transformative discovery. Once a brokerage connection is established between previously disparate areas, it would facilitate the information flow between these areas. In other words, we expect that the newly discovered connection will accelerate the diffusion process. Interestingly, the expected effect on diffusion can be explained in terms of the information foraging theory (Pirolli, 2007). According to the information foraging theory, searchers need to evaluate multiple patches of information. They need to make decisions on which patch they should focus and how long they should spend on a patch before they move on. Their decisions are essentially determined by a perceived profitability of each move. The higher the perceived profitability, the more likely they will decide to go ahead and take the action. The newly discovered connection will increase the perceived profitability because the discovery not only reduces the risk, but also provides concrete and positive examples of success. Therefore, we could conjecture that the increased perceived profitability will be translated into *bursts* of observed frequencies such as citation and co-citation counts.

Finally, the theory is related to but distinct from the notion of co-citation pathways through science (Small, 2000). The creation of co-citation pathways aims to traverse scientific literature through a chain of highly co-cited pairs of papers. Small found a co-citation pathway of 331 highly cited documents starting from economics and ending in astrophysics (Small, 1999). He observed that each successive document in this path embodies an information transition towards the destination topic and, in most cases, such transitions are surprisingly smooth. In contrast, the focus of our theory is on novel connections that bridge previously disparate fields. Although in theory such connections may appear as part of a co-citation pathway, it seems to be more likely that brokerage connections would either deviate from pathways of highly co-cited documents or not be selected altogether because of a high co-citation threshold. Nevertheless, more investigations are needed to clarify the relationships in detail.

### 3.2 Structural and Temporal Properties

Now we will focus on two specific properties of scientific discoveries that can be derived from our theory, namely a structural property of a discovery in a network of scientific papers measured by the betweenness centrality (Freeman, 1977) and a temporal property measured by burstness of citations (Kleinberg, 2002).

Our theory states that a transformative discovery is made when a bridging connection is established between two or more previously disconnected patches of knowledge. If we represent knowledge in the form of networks, such bridging connections would be links between two or more disconnected networks or components of a network. Such connections in networks can be computationally identified using the betweenness centrality. In fact, one can even compute the would-be centrality of a node if it were to have some of the non-existent connections. The betweenness centrality of a particular node or link measures the importance of the node or link in connecting any two nodes in the network. A node or link that is essential for linking many pairs of nodes will have a high betweenness centrality. Therefore, a paper with a high betweenness centrality is potentially a transformative discovery. In addition, it would be possible to use this metric to identify potential future discoveries by calculating the would-be betweenness centrality of a hypothetical connection between two disparate areas of existing knowledge networks. It is possible to devise computer simulation algorithms to identify a short list of such candidates of discovery to be made.

Several relevant concepts have been derived from betweenness centrality metrics. For example, in CiteSpace (Chen, 2004; Chen, 2006), pivotal points in co-citation networks are identified based on their betweenness centrality. These are the points that are cited with different co-citation clusters. We have mentioned earlier that co-citation clusters correspond to thematic structures. Therefore, points connecting different thematic structures are candidates of intellectual turning points.

In a journal co-citation network, high betweenness centrality is an indicator of interdisciplinary journals (Leydesdorff, 2007). Taken together, it suggests that the betweenness centrality indicator can be used at various scales of granularity to indicate and predict transformative changes. Furthermore, betweenness centrality is found to correlate with long-term citations predicted into the future (Shibata, Kajikawa, & Matsushima, 2007). This finding would be consistent with our conceptualization of scientific discovery in that scientists will pay constant attention to structural holes for future discoveries.

The emphasis on betweenness centrality differentiates our theory from other approaches to network evolution models, especially preferential attachment models. Instead of adding one link at a time to the most prominent node in a given network, our theory says that a scientific discovery needs to form a path spanning over an intellectual structural hole. As a result, the newly added scientific discovery would have a high betweenness centrality. Our theory also implies that a node with high betweenness centrality would be more valuable to a foraging scientist than a node with a higher citation count but lower betweenness centrality. While the latter may bring nothing new to a scientist who is well aware of the highly cited work, the former may lead to new insights that a scientist may actually act on. Thus, betweenness centrality can be translated into *interestingness,* which can be in turn translated into *actionability*. We have indeed observed in our previous work that the most cited references are not necessarily the most revolutionary ones (Chen, 2004; Chen & Kuljis, 2003).

Betweenness centrality is a structural property of a network. Our theory also leads to temporal properties of an evolving network, for example, the burstness of citation of a reference over time. Burst detection is a class of algorithms to identify changes of a variable over a period of time with reference to others in the same population (Kleinberg, 2002). Our theory suggests that a burst of citation could be a good indicator of a transformative discovery, especially from a profitability-guided foraging point of view, when it is observed with a structural property such as the betweenness centrality metric. As we have analyzed earlier, a brokerage discovery would increase the perceived profitability for moving from one patch of knowledge to another. As a result, the increased profitability and reduced risks should boost the adaptation and diffusion of the new discovery.

Would the absence of such structural and temporal properties rule out the possibility of a transformative discovery? This issue is concerned with the scope of the theory. However, we can only partially address this issue through some illustrative case studies in this article. Further investigations are needed. In the following section, we present some examples to further clarify the major properties derived from the theory.

## 3.3 Integration

In this article, we focus on cases in which both structural and temporal properties are observed and evaluate the role of brokerage mechanisms in such cases. In addition to study individual properties such as the betweenness centrality, the burst rate, or citation counts, we introduce a group of generic metrics $\boldsymbol{\sigma_n(v, G, T, \rho_1, \rho_2, \ldots, \rho_n)}$ as indicators of the potential transformative strength of a node $\boldsymbol{v}$ in a given network $\boldsymbol{G}$ over a time interval $\boldsymbol{T}$ with respect to $\boldsymbol{n}$ properties. Each $\boldsymbol{\rho_i}$ is a function $\boldsymbol{\rho_i}$ ($\boldsymbol{v, G, T}$) in the range of [0, 1]. These metrics can be generically defined as the geometric mean of multiple normalized properties $\boldsymbol{\rho_1, \rho_2, \ldots, \rho_n}$ in the range of [0, 1]. The maximum possible value of $\sigma$ is 1 when all the individual properties have the maximum value of 1. The minimum possible value of $\sigma$ is 0 when any of the individual properties is 0.

$$\sigma_n(v,G,T,\rho_1,\ldots,\rho_n) = \left(\prod_{i=1}^{n} \rho_i\right)^{\frac{1}{n}} \tag{1}$$

In particular, in the following case studies, the metric $\sigma$ is defined based on $\rho_{citation}$, $\rho_{centrality}$, $\rho_{burst}$ as follows. The definitions of $\rho_{centrality}$ and $\rho_{burst}$ can be found in (Brandes, 2001; Freeman, 1977; Kleinberg, 2002).

$$\sigma_1(v,G,T,\rho_{burst}) = \left(\prod_{i=burst} \rho_i\right)^{\frac{1}{1}} = \rho_{burst} \tag{2}$$

$$\sigma_1(v,G,T,\rho_{centrality}) = \left(\prod_{i=centrality} \rho_i\right)^{\frac{1}{1}} = \rho_{centrality} \tag{3}$$

$$\sigma_1(v,G,T,\rho_{citation}) = \left(\prod_{i=citation} \rho_i\right)^{\frac{1}{1}} = \rho_{citation} \tag{4}$$

$$\sigma_2(v,G,T,\rho_{burst},\rho_{centrality}) = \left(\prod_{i=burst,centrality} \rho_i\right)^{\frac{1}{2}} = \sqrt{\rho_{burst} \bullet \rho_{centrality}} \tag{5}$$

$$\sigma_3(v,G,T,\rho_{burst},\rho_{centrality},\rho_{citation}) = \left(\prod_{i=burst,centrality,citation} \rho_i\right)^{\frac{1}{3}} = \sqrt[3]{\rho_{burst} \bullet \rho_{centrality} \bullet \rho_{citation}} \tag{6}$$

Note that $\sigma_1(\rho_{citation})$, a special case of the generic definition, ranks the significance of a reference based on its citations as seen in earlier efforts for predicting Nobel Prize winners based on citation counts (Garfield, 1992). We will also compare pair-wise Pearson correlation coefficients between $\sigma_1$, $\sigma_2$ and $\sigma_3$ indices of centrality, burst, and citation frequency in order to identify the simplest and effective metrics among them.

In summary, our theory suggests that $\sigma$ indices would be a good indicator of potential transformative discoveries. Furthermore, once a reference is identified with a high $\sigma$ index, the theory provides an explanatory framework such that we can focus on the precise brokerage connections at work. The theory also suggests alternative ways to model the evolution of a network by taking brokerage connections into account. According to our theory, a subset of Nobel Prize discoveries will be transformative discoveries. More transformative discoveries would be expected from the recipients of a variety of other awards in science. In addition, we expect that transformative discoveries can be identified by these $\sigma$ metrics at an earlier stage than by single-dimensional ranking systems. In terms of diffusion, we expect that transformative discoveries in general will lead to a more rapid and sustained diffusion process. If we see the diffusion process as an information foraging process by the scientific community as a whole, transformative discoveries, i.e., brokerage connections across structural holes, would have a higher perceived profitability, which would motivate and stimulate the diffusion process. It also follows that the domain-wide foraging process will spend more time with transformative discoveries than other patches of scientific knowledge.

# 4 Illustrative Examples

We consider three examples as our initial verification of the theory. We choose two topic areas which have received Nobel Prize awards recently, namely, peptic ulcer and gene targeting, and string theory in physics as the third topic area.

## 4.1 Procedure

In each case study, CiteSpace (Chen, 2006) was used to construct a co-citation network of the references relevant to the chosen topic. We followed the general procedure described in (Chen, 2004; Chen, 2006). Bibliographic records were retrieved from the Web of Science with a topical search for articles only. Reviews, editorials, and other document types were excluded from the analysis.

CiteSpace uses a time-slicing mechanism to generate a synthesized panoramic network visualization based on a series of snapshots of the evolving network across consecutive time slices. Each node in the network represents a reference cited by records in the retrieved dataset. A line connecting two nodes represents one or more co-citation instances involving the two references. Colors of co-citation links correspond to the earliest year in which co-citation associations were first made. Each node is shown with a tree-ring of citation history in the same color scheme, representing the history of citations received by the underlying reference.

Structural-hole and burst properties are depicted in two distinct colors – purple and red – in visualizations. If a node is rendered with a purple ring, it means it has a strong betweenness centrality. The purple color can only appear as the color of the outermost rim of a node. The thickness of the purple ring is proportional to the degree of the centrality: the thicker, the stronger the betweenness centrality. In contrast, if a node has red rings, these red rings represent the presence and strength of its burst property. It can appear as the color of any inner rings of the tree ring of a node. The presence of one or more red rings on a node indicates a significant citation burst was detected. In other words, there was a period of time in which citations to the reference increased sharply with respect to other references in the pool, hence the name CiteSpace.

## 4.2 Case Study I: Peptic Ulcer

The Nobel Prize in Physiology or Medicine for 2005 was awarded jointly to Barry J. Marshall and J. Robin Warren for their discovery of "the bacterium *Helicobacter pylori* and its role in gastritis and peptic ulcer disease." We choose *peptic ulcer* as the topic area.

According to Marshall's Nobel Prize lecture (Barry J. Marshall, 2005), Marshall and Warren conducted a study in the 1980s and found 100% of 13 patients with duodenal ulcer were infected by Helicobacter pylori. They discovered that peptic ulcer was caused by a bacterial infection, unlike the then predominant understanding that ulcers were caused by other reasons such as stress and acid in the stomach. The discovery established that very young children acquired the Helicobacter organism, a chronic infection which caused a lifelong susceptibility to peptic ulcers. Helicobacter was generally accepted after 1994 as the cause of most gastroduodenal diseases including peptic ulcer and gastric cancer.

We analyzed a co-citation network of peptic ulcer research to identify structural and temporal properties associated with the Helicobacter pylori discovery. Bibliographic records on peptic ulcer between 1980 and 2007 were retrieved from the Web of Science with a topic search for 'peptic ulcer.' CiteSpace was used to construct a co-citation network of peptic ulcer research between 1980 and 2007.

Figure 1 shows a series of 5-year snapshots of the co-citation network as it evolved over time. In each diagram, five colors match to the five years in the order of blue, cyan, green, yellow, and orange. Thus, an orange cluster would be formed in the 5$^{th}$ year of a given 5-year interval. For example, a node with essentially a green tree-ring means the reference was mostly cited in the 3$^{rd}$ year of the time interval.

The captions below network snapshots record the time interval, the number of nodes, the number of co-citation links, and three thresholds. For example, the caption "1981-1985. N=210, E=2038. 3,3,20" under the first snapshot of the network means that the network was formed between 1981 and 1985, consisting of 210 references and 2,038 co-citation pairs. Each reference has received at least 3 citations in one of the 5 years during this period.

According to independent sources (Pincock, 2005), the first major publication of the Helicobacter pylori discovery was (B. J. Marshall & Warren, 1984). Marshall-1984 appeared in the 1986-1990 network with essentially cyan and green citation rings, which means it received its citations mostly in 1987 and 1988. It is quite possible that Marshall-1984 was cited as soon as it was published in the 1981-1985 time interval, but it did not reach the top of the most cited list until the 1986-1990 network. The six snapshots also demonstrate that peptic ulcer research has evolved constantly with new references reaching the top cited levels.

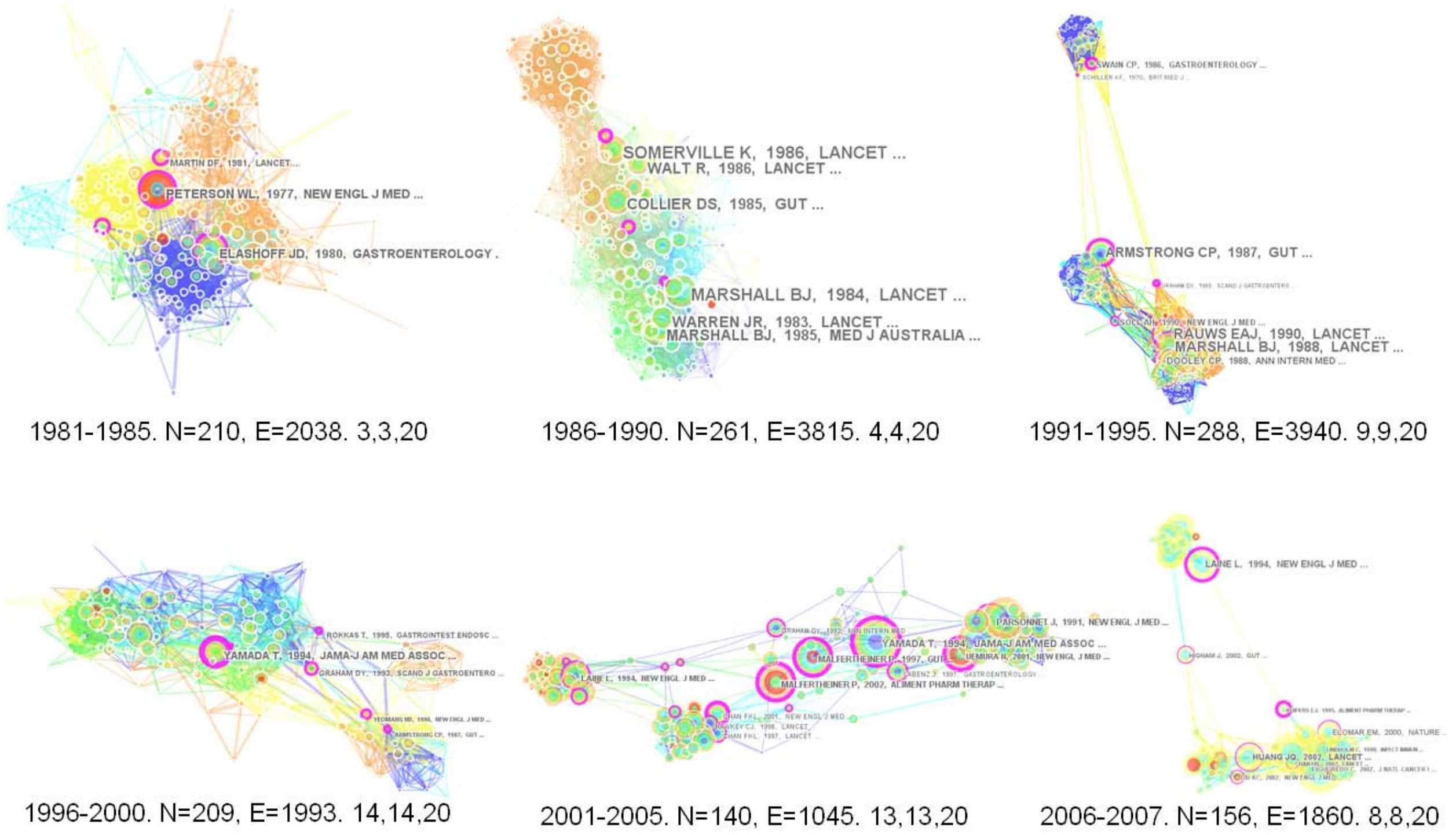

**Figure 1. A co-citation network of references on peptic ulcer research (1980-1990).**

Figure 2 shows a panorama view of the entire time interval of the dataset (1980-2007). Marshall-1984 has a prominent structural property – a high betweenness centrality (a large purple ring). Although it does demonstrate a temporal property of burstness, its burst rate is detectable but not as strong as some of its neighbors. The burst period was between 1986 and 1988, which is consistent with our observations in the earlier 5-year snapshot series. The overview network shows that Marshall-1984 is in a dense cluster with numerous references with citation bursts, suggesting other high-impact references were present in the landscape of peptic ulcer research.

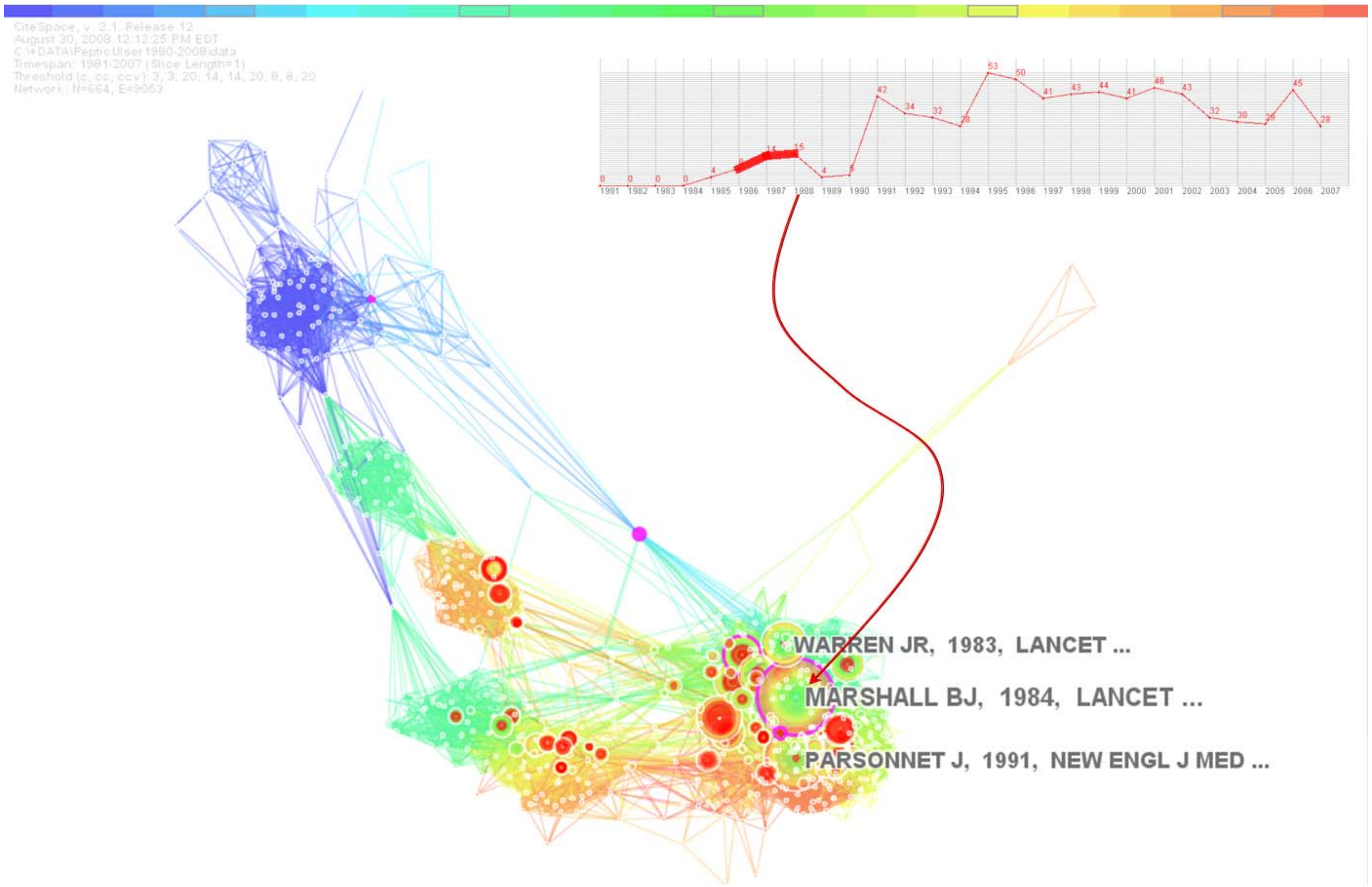

**Figure 2. A co-citation network of references cited between 1981 and 2007 in peptic ulcer research.**

As shown in Table 1, Marshall-1984 was the most cited reference (711 citations) and the highest betweenness centrality ($\rho_{centrality}$ of 0.393). On the other hand, its burst rate ranked the 372$^{nd}$. Marshall and Warren encountered resistances in getting their discovery accepted by the peptic ulcer research community. The slow acceptance was documented (Pincock, 2005), which may in part explain its relatively low burst rate. In contrast, Marshall-1988 has the highest $\sigma_2$ of 0.416. It was entitled *Prospective double-blind trial of duodenal ulcer relapse after eradication of Campylobacter pylori.* In his Nobel Prize lecture, Marshall dated the acceptance of his work as the 1994 NIN consensus conference in Washington DC.

**Table 1. Top 5 most cited references in peptic ulcer research (1980-2007).**

| Citation | Author | Year | Source | Vol. | Page | $\rho_{burst}$ | $\rho_{centrality}$ | $\sigma_2$ |
|---|---|---|---|---|---|---|---|---|
| **711** | MARSHALL BJ | 1984 | LANCET | 1 | 1311 | 0.138 | 0.393 | 0.232 |
| **581** | PARSONNET J | 1991 | NEW ENGL J MED | 325 | 1127 | 0.208 | 0.143 | 0.172 |
| **579** | WARREN JR | 1983 | LANCET | 1 | 1273 | 0.165 | 0.250 | 0.203 |
| **466** | YAMADA T | 1994 | JAMA | 272 | 65 | 0.635 | 0.071 | 0.213 |
| **421** | MARSHALL BJ | 1988 | LANCET | 2 | 1437 | 0.607 | 0.286 | 0.416 |

The last column in Table 1 contains the $\sigma_2$ index, i.e., the geometric mean of the burst and centrality metrics. According to our theory, a transformative discovery is a brokerage between previously disconnected areas of scientific knowledge. The $\sigma_2$ index takes into account both structural and temporal properties that a discovery over a structural hole would demonstrate. In this case, Marshall-1988 was the highest ranking candidate according to the $\sigma_2$ index, despite its citation count of 421 was much less than Marshall-1984. Validating the true value of Marshall-1988 is beyond our own expertise and beyond the scope of the article. Properly validating the value of references with such strong combinations of structural and temporal properties will be an important issue to be addressed in the future work of our construction of the theory. It is also related to the potential power of predicting high-impact discoveries even before it reaches its citation peaks or while they are overshadowed by other highly cited references.

### 4.3 Case Study II: Gene Targeting

The Nobel Prize in Physiology or Medicine for 2007 was awarded jointly to Mario R. Capecchi, Martin J. Evans and Oliver Smithies for their discoveries of "*principles for introducing specific gene modifications in mice by the use of embryonic stem cells.*" This field of study is often known as gene targeting. We applied the same procedure described earlier for gene targeting. We used topic searches in the Web of Science for ‘gene target*’, ‘genetic* target*’, and ‘gene* knock*’ for genetic knock-out, another term used to describe the techniques in general. A total of 8,160 bibliographic records were retrieved between 1985 and 2007.

Figure 3 shows an overview of a co-citation network of gene targeting references cited between 1985 and 2007. Notably, the three nodes with the highest betweenness centrality scores are all connected to the 2007 Nobel Prize awards: Capecchi-1989, Mansour-1988, and Thomas-1987. Here only the first author of each paper was recorded in the Web of Science cited reference field. The three papers represent a series of innovations of fundamental techniques for gene targeting. Unlike the case with Marshall-1984, all three groundbreaking gene targeting papers have strong citation bursts, shown in Figure 1 as the thickened rising curves. It also becomes clear that these curves have subsequently peaked and steadily declined, which means they are getting fewer and fewer citations. The visualization confirms this pattern. The network shows that the most recent active areas are located in the lower left quadrant of the visualization.

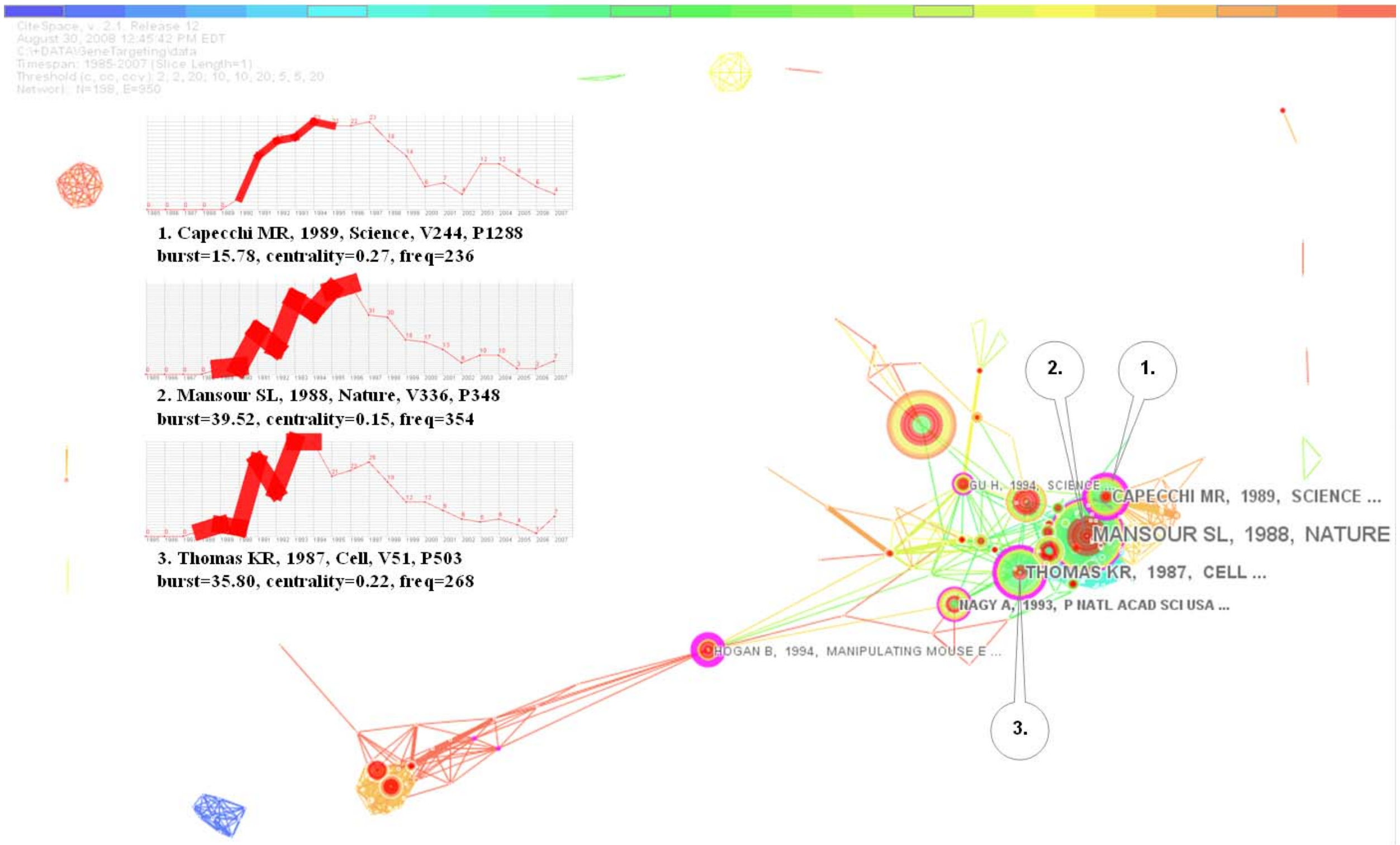

**Figure 3. A co-citation network of references cited between 1985 and 2007 in gene targeting research. References with the strongest betweenness centrality scores are labeled. The burst periods of their citations are shown as the thickened curves in the three diagrams to the left.**

The 2007 Nobel Prize awards mentioned the use of *embryonic stem cells*. Techniques developed in embryonic stem cell research turned out to be critical to the gene targeting techniques. Martin J. Evans, who shared the 2007 Nobel Prize, is known as the architect of embryonic stem cells. The pioneering discovery made by Evans in 1981 (Evans & Kaufman, 1981) was in fact cited by the Thomas-1987 gene targeting paper. Evans-1981 was cited 1,681 times in the Web of Science, although it was not highly cited within the gene targeting dataset we analyzed. Techniques developed by Evans were among the many building blocks that were necessary for the ultimate gene targeting techniques. A number of questions can be addressed from our theory of discovery. For example, how easy or how hard was it to discover Evans-1981 for the needs of gene targeting? Who were the first citers of Evans-1981. What was Evans' own research field and how was it related to gene targeting? What are the other building blocks used by these Nobel laureates in their discoveries? Were their discoveries taking place over an intellectual structural hole? How did their discovery change the association between existing intellectual structures?

Table 2 lists the top 5 references by $\sigma_2$ – the geometric mean of $\rho_{centrality}$ and $\rho_{burst}$. The 1st, 3rd, and 4th references are connected to the Nobel Prize winning discoveries. Note that the first discovery paper Thomas-1987 has the highest ranking although its citation count of 268 is not the highest. The 2nd reference is a book. If we consider journal articles only, the first three references would be all related to the Nobel discoveries.

**Table 2. Top 5 references by $\sigma_2$ – the geometric mean of centrality and burstness.**

| Author | Year | Source | Vol. | Page | Citations | $\rho_{burst}$ | $\rho_{centrality}$ | $\sigma_2$ |
|---|---|---|---|---|---|---|---|---|
| THOMAS KR | 1987 | CELL | 51 | 503 | 268 | 0.851 | 0.537 | **0.676** |
| HOGAN B | 1994 | MANIPULATING MOUSE E | BOOK | | 136 | 0.409 | 1.000 | **0.639** |
| MANSOUR SL | 1988 | NATURE | 336 | 348 | 354 | 0.940 | 0.366 | **0.586** |
| CAPECCHI MR | 1989 | SCIENCE | 244 | 1288 | 236 | 0.375 | 0.659 | **0.497** |
| NAGY A | 1993 | P NATL ACAD SCI USA | 90 | 8424 | 182 | 0.346 | 0.463 | **0.400** |

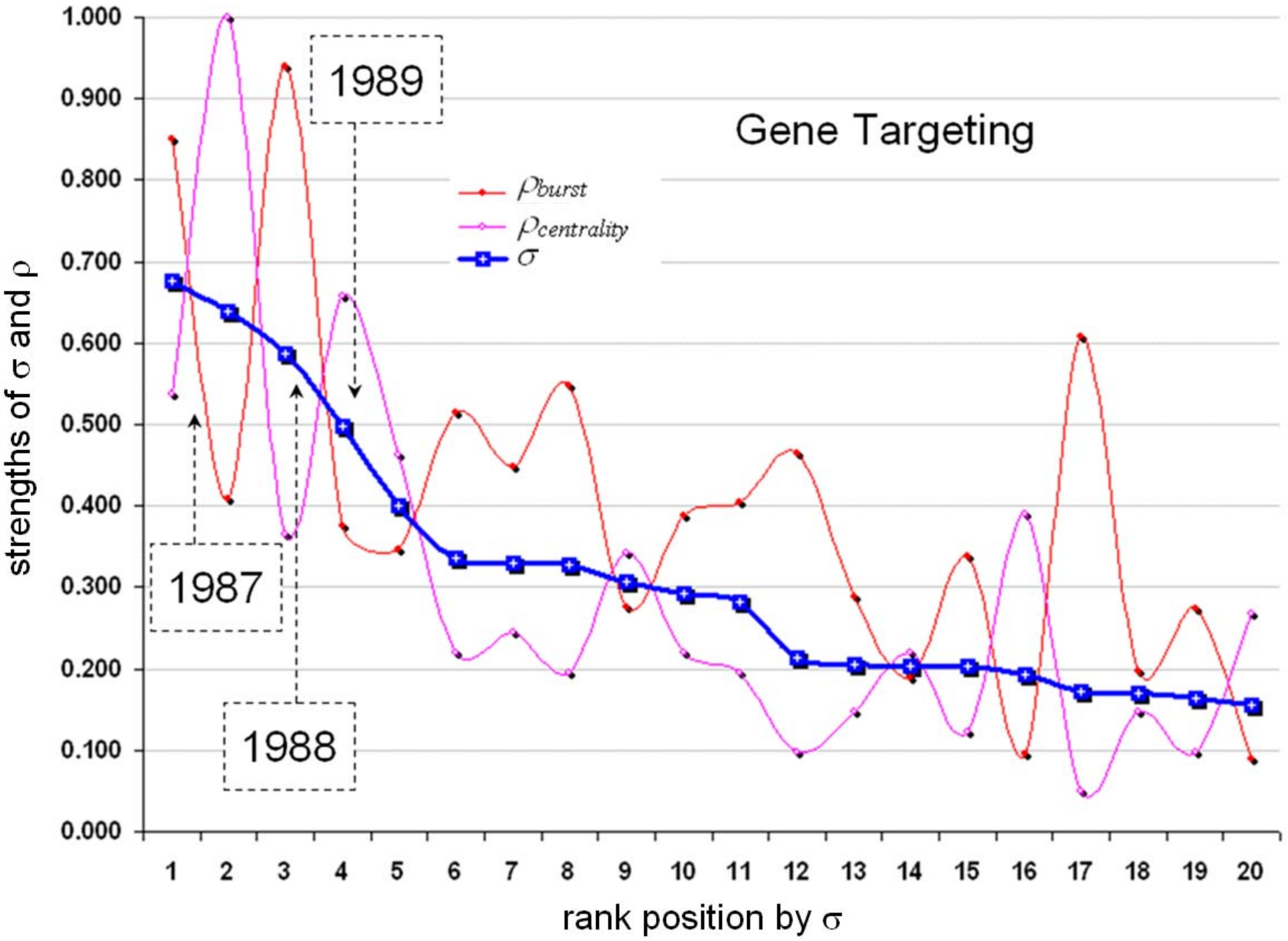

**Figure 4. Nobel Prize winning discovery papers are ranked among the highest by the $\sigma_2$ index.**

Figure 5 is a visualization of the areas associated with the Nobel Prize winning discoveries in gene targeting research. The visualization was generated based on citing articles with 15 or more citations in the Web of Science. In other words, these citing articles themselves have made impacts on the field in their own right. Co-cited references are aggregated into clusters. The diffusion of knowledge is tracked by showing how co-citation footprints move from one cluster to another over time and how long they stay in particular clusters. The history of the evolution can be seen as an information foraging process participated in by all the scientists in the field. For example, the *embryo-derived stem cell* (cluster #11) attracted a lot of citations in 1987 (shown as a high density cocitation cluster in red). In 1988, the foraging process moved to *DNA delivery method* (cluster #19) above cluster #11. All three papers associated with the 2007 Nobel Prize are concentrated in cluster #12 – *gene correction*. During 1989 and 1990, much of the foraging process was inside cluster #12. We also studied the diffusion process over a longer period of time and the foraging process appeared to spend much longer time with cluster #12 than any other clusters. Our general hypothesis is that transformative discoveries tend to retain the foraging process longer than other patches of knowledge. Further investigations are needed.

The connection between structural-hole theory and information foraging theory is an important research direction for further investigation.

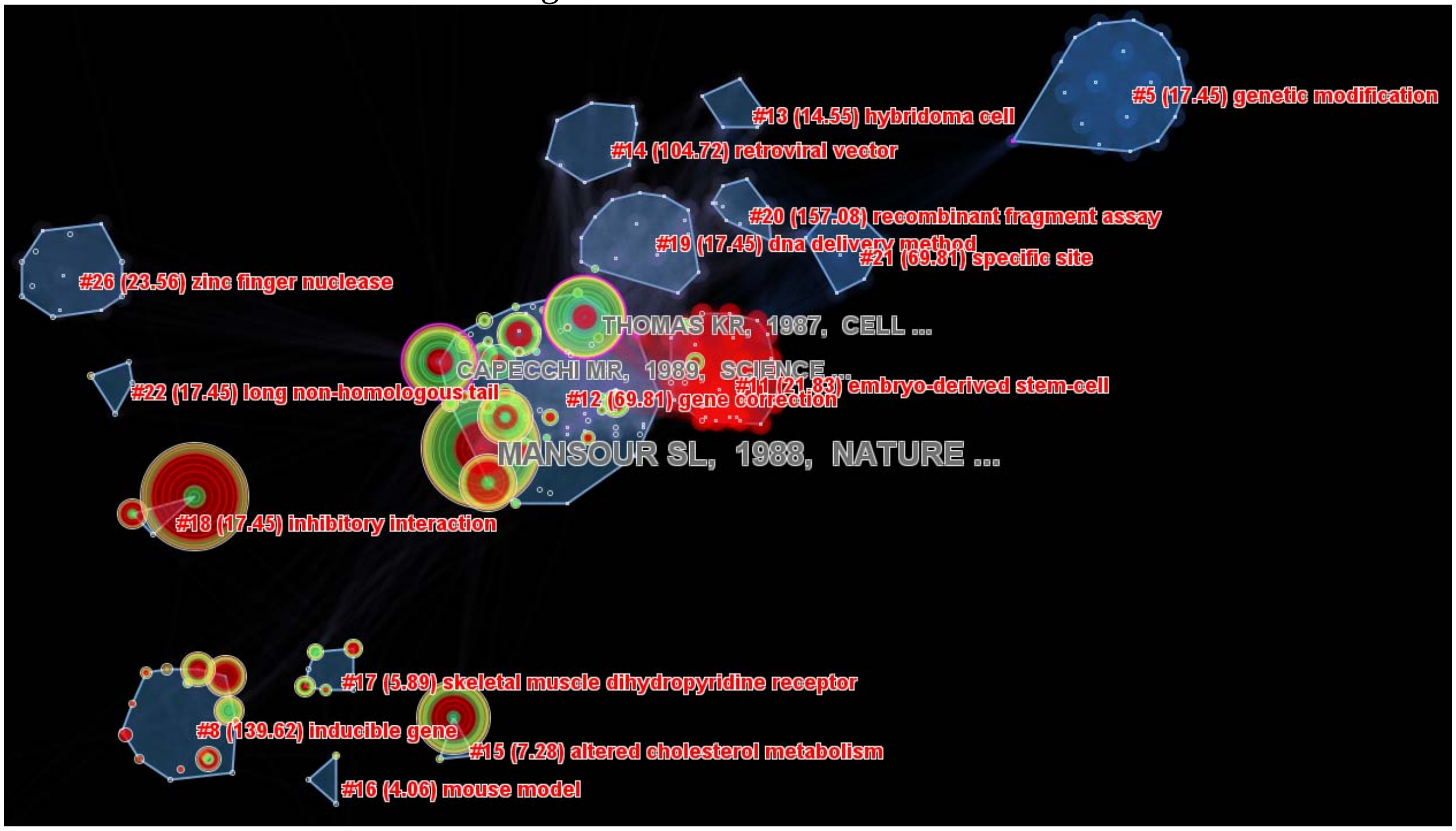

**Figure 5. A diffusion map of gene targeting research between 1985 and 2007. Selection criteria are at least 15 citations for citing articles and top 30 cited articles per time slice. Polygons represent clusters of co-cited papers. Each cluster is labeled by title phrases selected from papers citing the cluster. Red lines depict co-citations made in the current year. The concentrations of red lines track the context in which co-citation clusters are referenced.**

## 4.4 Case Study III: String Theory

The third illustrative example is string theory in physics (Schwarz, 1982). We have studied this topic as an example of Kuhn's scientific revolutions (Chen, 2004; Chen & Kuljis, 2003). According to (Schwarz, 1982), two conceptual revolutions occurred in string theory: one was in 1980s and the other in 1990s. Using bibliographic records published between 1990 and 2003, we conducted a similar study of string theory and focused on the two properties of the revolutionary paper for the second string theory revolution.

Figure 6 shows an overview of a visualized co-citation network of references in the period of 1990-2003. According to Schwarz (1982), Polchinski-1995 marked the second string theory revolution. Polchinkski-1995 is ranked the 5$^{th}$ by the geometric mean index. The visualization shows it has a relatively strong betweenness centrality and its burst rate is not as prominent as a few others in the field. Witten-1991 has the highest geometric mean index ranking, followed by Maldacena-1998; both have shown strong betweenness centrality and burstness.

Maldacena-1998 is not only strong in both centrality and burstness, it is also the most cited reference in this dataset. We contacted Juan Maldacena directly and asked him to identify the nature of his major contributions in this article to String Theory. The transformative nature is evident in his reply: "It connected two different kinds of theories: 1) particle theories or gauge theories and 2) string theory. Many of the papers on string dualities (and this is one of them) connect different theories. This one connects string theory to more conventional particle theories." Maldacena's contribution is highlighted on the TIME 100 Innovator website[2] as "he forged a connection between the esoteric formulas of string theory and the rest of mainstream physics." Even more intriguingly from the perspective of our brokerage theory, he "has been able to suggest a way to knit together two theories previously thought to be incompatible: quantum mechanics, which deals with the universe at its smallest scales; and Einstein's general theory of relativity, which deals with the very largest." In addition, our search on the web reveals that he is the recipient of the 2007 Dannie Heineman Prize for Mathematical Physics[3] "for profound developments in Mathematical Physics that have illuminated interconnections and launched major research areas in Quantum Field Theory, String Theory, and Gravity."

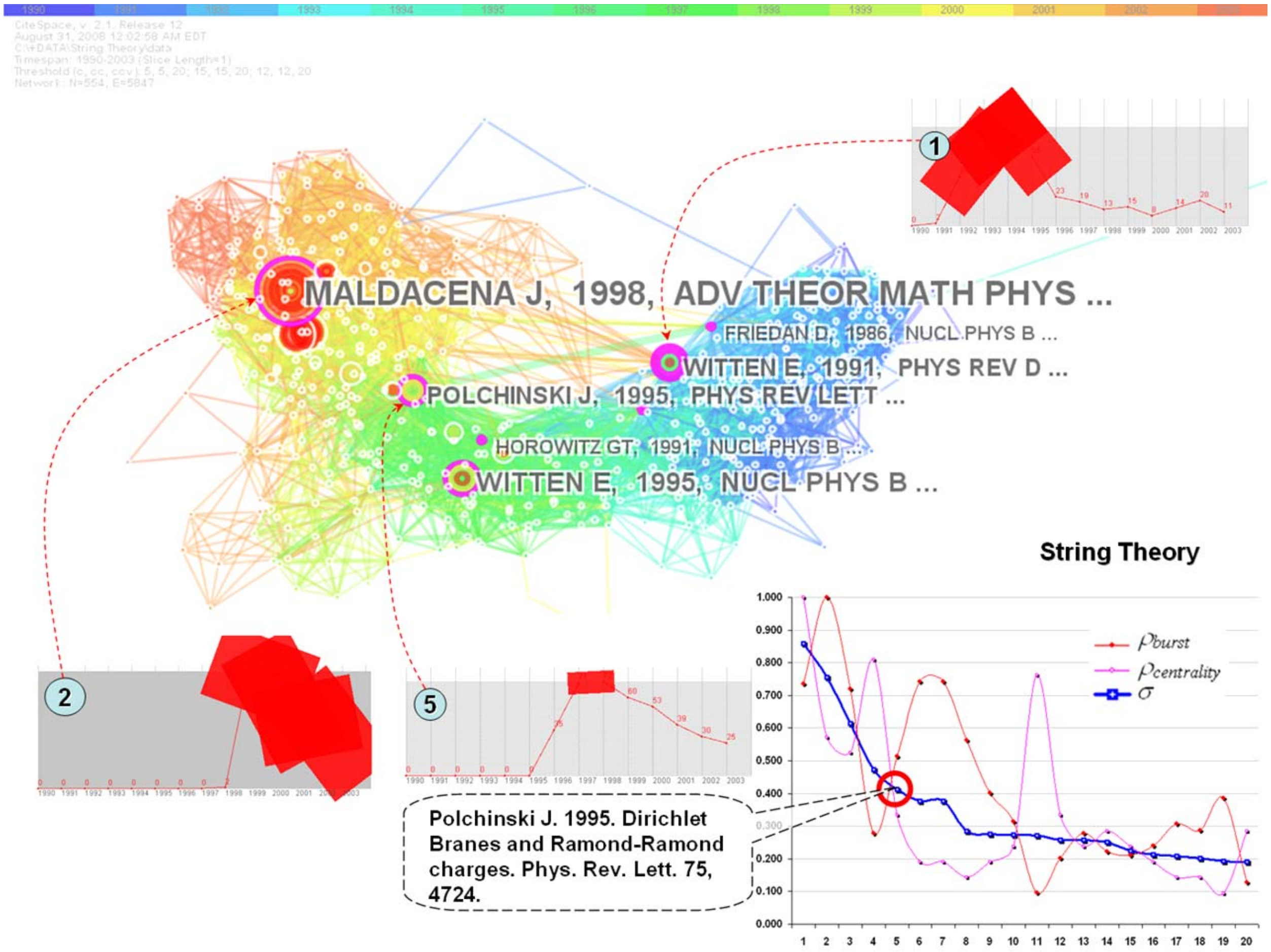

**Figure 6. A co-citation network of references cited between 1990 and 2003 in string theory. Polchinski-1995 marked the beginning of the second string theory revolution. Maldacena-1998 is highly transformative and brokerage link between string theory and particle theories. The three embedded plots show the burst periods of citations of Witten-1991, Maldacena-1998, and Polchinski-1995.**

Table 3 shows pair-wise Pearson correlation coefficients between normalized burst and centrality scores, the $\sigma_2$ index of burst and centrality, and the $\sigma_3$ index of burst, centrality, and citation frequency. The $\sigma_2$ and $\sigma_3$ indices are strongly correlated (r=0.9780), suggesting that, at least in this case, the $\sigma_3$ index is redundant and we can simply focus on $\sigma_2$. The correlation coefficients also show that burstness and centrality are almost independent measures, although they both have some connections to citation counts. This is a simple justification of our choice to use both burstness and centrality to construct $\sigma_2$ as an index of high-impact discoveries. More comprehensive validations may consider other measures such as the h-index and its numerous variations, e.g. (Antonakis & Lalive, 2008; Hirsch, 2005b).

[2] http://www.time.com/time/innovators/science/profile_maldacena.html
[3] http://www.aps.org/programs/honors/prizes/prizerecipient.cfm?name=Juan%20Maldacena&year=2007

**Table 3. Pearson correlation coefficients between individual properties and synthetic indices.**

| | $\rho_{burst}$ | $\rho_{centrality}$ | $\sigma_2(\rho_{burst}, \rho_{centrality})$ |
|---|---|---|---|
| $\rho_{citation}$ | 0.8026 | 0.3618 | |
| $\rho_{burst}$ | | 0.0409 | |
| $\sigma_3(\rho_{burst}, \rho_{centrality}, \rho_{citation})$ | | | 0.9780 |

# 5 Discussions and Conclusions

We have introduced an explanatory and computational theory of transformative discovery in science. The theory focuses on the role of making connections across structural holes between two or more network representations of scientific knowledge in scientific discovery. This theory provides a conceptual framework to connect a diverse range of theories of scientific change, social capital of structural holes, and an extended information foraging theory for knowledge diffusion.

## 5.1 Major Contributions

The theory lends itself to multiple instantiations of philosophical theories, sociological theories, and information theories of intellectual change. For example, the emergence of new paradigms can be seen as a newly established connection between existing disciplines. We know it is often the case that scientific discoveries draw inspiration and critical enabling techniques from areas outside their home fields as well their own ones. Paradigm-shifting changes would be detectable as the center of citations moves in terms of structural properties' change. From a sociological perspective, the competition driven scientific discoveries can be explained in terms of the competitive advantages brought in by the generalized notion of structural holes. In addition, information theories and information foraging theories can explain why scientists should pay special attention to interdisciplinary structural holes to maximize the profitability of their moves. Betweenness centrality, citation burstness and the proposed metric can be used to provide information scents along an information foraging path through a bibliographic forest. With the recognition of the significance of these metrics a forager could make better estimates of the profitability of alternative pathways.

The theory suggests that a high-impact discovery should be strong in both structural and temporal properties: namely, the betweenness centrality and citation burstness. This theory is built on the structure and dynamics of networks. It can be seen as an expansion of theories and growth mechanisms based on citation counts alone, such as citation frequencies and h-index. The three illustrative examples have shown that the two properties characterize our cases reasonably well. The peptic ulcer case is an example in which citation burstness was relatively low. It may be the case for a wider range of scientific change that has to fight for its acceptance. In contrast, the gene targeting case is a satisfying example in which both properties are strongly related to Nobel Prize winning discoveries. The string theory case was able to identify the paper that trigged the second string theory revolution among a short list of top 5.

In a relevant study, also in this special issue, Bettencourt et al. focused on the changes in the structure of collaboration of an emerging field (Bettencourt, Kaiser, & Kaur, This Issue). They conjectured that new conceptual breakthroughs would typically lead to not only a rapid growth in the number of scientists and publications, but also an increasingly tighter collaboration among scientists. Indeed, our theory may help to identify mechanisms that lead to the increase of the density of collaboration after a transformative scientific discovery. For example, the brokerage nature of a transformative discovery implies that it will change the perceived return-risk ratio, which in turn fosters the diffusion of knowledge, as a social navigation and foraging process, across previously disparate areas of research. In other words, a newly established conceptual pathway would open up new ways for scientists to collaborate.

The geometric mean of betweenness centrality and burstness is the first index derived from the theory of scientific discovery. It identifies high-impact original discoveries and partially overcomes the scenarios in which original publications were overshadowed by other highly cited references. On the other hand, the three case studies also revealed that the discovery index may identify additional publications. The status of such publications should be thoroughly investigated. Are they in the same status as the Nobel Prize worthy discoveries? If so, it will be an encouraging means to identify such discoveries ahead of Nobel Prize awards. If, on the other hand, the prominence of these references was due to other reasons, one should identify these reasons and use these reasons to mark the scope of the theory of discovery. For example, is Marshall-1988 more significant than Marshall-1984? Why is Witten-1991 so prominent in both properties in string theory?

Such unanswered questions may also provide potential research targets for historians of science, philosophers and sociologists of science. For computer and information scientists, the proposed theory offers a new framework for simulating the growth and decay of intellectual networks. One may compare and combine different growth mechanisms such as single-link preferential attachment and trailblazing mechanisms over intellectual and disciplinary structural holes. If we define a semantic metric that measures the intellectual distance between both sides of a structural hole, the theory implies that the larger the gap, the higher the potential impact. In terms of the expanded information foraging theory, the expected returns are only part of the equation that foragers have to consider. The risks can be reduced by spreading out the search and maintaining a relatively low-cost of weak-ties with scientists who are different from ourselves, as observed by Crane, Burt, and others.

## 5.2 Limitations and Future Work

The study of scientific discovery has a very broad, complex, and multidisciplinary scope. There are many other approaches that we have not covered in this article, for example, modeling evolving networks (Ausloos & Lambiotte, 2007; Bruckner, Ebeling, & Scharnhorst, 1990; Koenig, Battiston, & Schweitzer, 2008; Lambiotte & Ausloos, 2007), citation histories of scientific publications (Vlachy, 1985), Nobel Prize discoveries (Czerwon & Vlachy, 1986; Zuckerman, 1967), and formal models of scientific revolutions (Sterman, 1985).

As open-notebook science and e-science increase in popularity more of the data generated during the discovery process is becoming accessible to researchers. This data leaves additional breadcrumbs for tracking the paths of innovation. Such data could provide another view into the discovery process. The case studies described here rely on bibliographic data as an indicator of the connectedness of ideas. New metrics might be developed from other sources such as the contents of open laboratory notebooks. These sources could be useful for the creation of additional metrics to supplement bibliometric analyses. Open notebook data would be particularly valuable for studying the structure of paths which do not lead to Nobel Prize winning discoveries but do constitute the bulk of scientific inquiry.

The work described in this article is the first step in our ongoing search towards a better understanding of scientific discovery. Much work remains to be done to validate the theory with large samples of discoveries and identify the status of various references highly ranked by the geometric mean index of high-impact discovery. Many issues concerning the acceptance and rejection of a new discovery need to be addressed. For example, why do we find some ideas interesting whilst remaining indifferent to other ideas or strongly reject them? What is the potential connection to interestingness research in the data mining community (Hilderman & Hamilton, 2001; Silberschatz & Tuzhilin, 1996; Tan, Kumar, & Srivastava, 2002)? How do we deal with uncertainties in puzzle solving and mystery solving processes (Chen, 2008)? How is the theory quantitatively related to existing models of knowledge diffusion, such as epidemic models, ant colony, and random walk models? How should the theory connect to the findings in literature-based discovery? Should we differentiate use-driven discoveries from other discoveries outside the so-called Pasteur's quadrant (Stokes, 1997)? If so, what properties are useful?

Our long-term goal is to contribute to the understanding of generic mechanisms for scientific discovery and the dynamics of scientific fields. The theory underlines the value of interdisciplinary collaboration in science and the diffusion of knowledge. Given the increasing interests in cyber-enabled discovery, e-science, and e-social science, the proposed theory is expected to serve as a starting point for integrating conceptualizations of scientific change from multiple disciplines and for empirical studies of science.

**Acknowledgements**

This work is supported in part by the National Science Foundation (NSF) under grant number 0612129, the US Department of Homeland Security through the Northeast Visualization and Analytics Center (NEVAC), and the Chang Jiang Scholar program of the Chinese Ministry of Education. Thanks to Xianwen Wang and Xiaoyu Zhu of Dalian University of Technology, China, for their assistance in data collection and preliminary analysis of gene targeting and peptic ulcer cases, Juan Maldacena, Institute for Advanced Study, for responding to our inquiries, and anonymous reviewers for their detailed and constructive comments.

**Notes**

CiteSpace is freely available at http://cluster.cis.drexel.edu/~cchen/citespace. Color versions of the figures in this article are available at http://cluster.cis.drexel.edu/~cchen/papers/JOI.